\documentclass[journal]{IEEEtran}
\usepackage{cite}
\ifCLASSINFOpdf
  \usepackage[pdftex]{graphicx}
  \DeclareGraphicsExtensions{.pdf,.jpeg,.png}
\else
  \usepackage[dvips,xetex]{graphicx}
  \DeclareGraphicsExtensions{.eps,.pdf,.jpeg,.png}
\fi
\graphicspath{{./fig/}}
\usepackage{amsmath}
\usepackage{autobreak}
\usepackage{amsthm}
\usepackage{amssymb}
\usepackage{bbm}
\usepackage[ruled]{algorithm2e}
\usepackage{array}
\usepackage{cleveref}
\ifCLASSOPTIONcompsoc
  \usepackage[caption=false,font=normalsize,labelfont=sf,textfont=sf]{subfig}
\else
  \usepackage[caption=false,font=footnotesize]{subfig}
\fi
\usepackage{threeparttable}
\usepackage{multirow}
\usepackage{multicol}
\usepackage{booktabs}
\usepackage{url}
\begin{document}
\title{Two-stage Deep Reinforcement Learning for Inverter-based Volt-VAR Control in Active Distribution Networks}
\author{Haotian~Liu, ~\IEEEmembership{Student~Member,~IEEE},
  Wenchuan~Wu,~\IEEEmembership{Senior~Member,~IEEE}
  \thanks{This work was supported in part by the National Natural Science Foundation of China (Grant 51725703).}
  \thanks{H. Liu and W. Wu (Corresponding Author)  are with the State Key Laboratory of Power Systems, Department of Electrical Engineering, Tsinghua University, Beijing 100084, China (email:lht18@mails.tsinghua.edu.cn, wuwench@tsinghua.edu.cn).}
}
\markboth{Journal of \LaTeX\ Class Files,~Vol.~xx, No.~x, August~20xx}
{Shell \MakeLowercase{\textit{et al.}}: Bare Demo of IEEEtran.cls for IEEE Journals}
\maketitle

\begin{abstract}
  Model-based Vol/VAR optimization method is widely used to eliminate voltage violations and reduce network losses.
  However, the parameters of active distribution networks(ADNs) are not onsite identified, so significant errors may be involved in the model and make the model-based method infeasible.
  To cope with this critical issue, we propose a novel two-stage deep reinforcement learning (DRL) method to improve the voltage profile by regulating inverter-based energy resources, which consists of offline stage and online stage.
  In the offline stage, a highly efficient adversarial reinforcement learning algorithm is developed to train an offline agent robust to the model mismatch.
  In the sequential online stage, we transfer the offline agent safely as the online agent to perform continuous learning and controlling online with significantly improved safety and efficiency.
  Numerical simulations on IEEE test cases not only demonstrate that the proposed adversarial reinforcement learning algorithm outperforms the state-of-art algorithm, but also show that our proposed two-stage method achieves much better performance than the existing DRL based methods in the online application.
\end{abstract}

\IEEEpeerreviewmaketitle

\section{Introduction}
\IEEEPARstart{V}{olt/VAR} control (VVC) has been successfully integrated into distribution management system to optimize the reactive power flow, achieving the goal of eliminating the voltage violations and reducing network losses.  Conventionally, VVC is a model-based optimization method to generate  a set of optimal strategies for voltage regulation devices and reactive power resources\cite{6213584}.

With an increasing penetration of distributed generations (DG) in active distribution network (ADN), the problems of voltage violations and high network losses are becoming more severe, especially in the case of reversed active power flow\cite{6336354,8960272}. Due to the fact that most DGs are inverter-based energy resources (IB-ER) and typically produce less active power than the rated capacity, it is reasonable and required for the IB-ERs to provide Volt/VAR support.

Till now, most VVC methods are model based. This is to say, the problem of VVC has been described as a nonlinear programming problem.
The majority of the existing VVC algorithms in both centralized manners or decentralized manners employ various optimization techniques with real-time measurements, which rely on the accurate model of the physical system.
The common centralized VVC algorithms include the well-known conic relaxation methods, interior point methods \cite{4808225}, mixed integer linear programming\cite{BORGHETTI201339}, and evolutionary algorithms \cite{6336354, 5649862}. Also, plenty of literatures adopted decentralized algorithms under different structures, including local control \cite{7361761}, quasi real-time reactive optimization\cite{7470528}, alternating direction method of multipliers (ADMM) \cite{7926415,7042735} and accelerate ADMM based distributed control \cite{8960272}. To address the uncertainty issues regarding DERs, robust optimization \cite{zwy2017} and scenario based uncertainty optimization \cite{8419280} are also proposed.

However, in ADNs, the network parameters are theoretical parameters instead of onsite identified ones, which cannot reflect the real operation states and significant errors are involved \cite{6515297, 6579907, 7350258, 8944292}.
The model mismatch issue hinders the application of existing model-based methods in the real world. 
Therefore, model-free control is an alternative and promising solution for VVC, which learns optimal actions with only measurements data and continuous exploration in the action space.
Legacy model-free optimization methods were mainly applied to the wind farm control using game theory \cite{6515297,haarnoja2018sac,haarnoja2018sacapplications} or gradient-based optimization\cite{6579907}.
In recent years, deep reinforcement learning (DRL) algorithms, which are able to work in a model-free manner, have demonstrated remarkable performance on multiple controlling tasks such as games\cite{mnih2015human,silver2016mastering}, autonomous driving\cite{shalev2016safe, pan2017virtual} and continuous control\cite{duan2016benchmarking,lillicrap2015continuous}.
Hence, DRL-based VVC algorithms have been developed and compared to traditional optimization-based methods in \cite{li2019coordination, 8909741}.
In these works, DRL-based VVC methods have shown notable improvement on the performance.

However, the online training start-up of the agents in DRL-based VVC methods can lead to unbearable cost and high risk, since the algorithm has little knowledge of the real system. It is reasonable and desirable for the agents to implement offline training before online training and application. 
Since the configuration of ADNs are continually updated, historical data is not appropriated for offline training in many scenarios. Therefore, an ADN simulation model is needed to train the agent offline. As mentioned above, the parameters of ADN are inaccurate, so special learning algorithm is indispensable to accommodate this situation. The agents trained on an inaccurate model may show undesirable performance when applied to the real system, which is called the ``transfer gap''.

In this paper, we propose a two-stage deep reinforcement learning method to optimize the reactive power distribution in ADNs, which including offline learning stage and online learning \& application stage.
Firstly, we build an approximate ADN model using theoretical parameters. Then, the offline trained agent based on such model are transferred to online stage to improve the online exploration efficiency and safety. Specifically, we formulate the VVC problem as Markov decision process (MDP) in the online stage.

In the offline stage, we propose a novel joint adversarial soft actor-critic (JASAC) algorithm with the formulation of adversarial MDP (AMDP) and view the modeling error in the simulation as an extra disturbance to the system. An adversary agent on behalf of the modeling error and risk is utilized to make the main (i.e., protagonist) agent robust to the model mismatch, which is called the offline agent (OFF-A). JASAC guarantees an efficient convergence of the adversarial training process taking advantage of the shared information between the adversary agent and OFF-A. In the online stage, we implement the online agent (ON-A) with knowledge of OFF-A using the state-of-art DRL algorithm called soft actor-critic (SAC). Thus, the online training process is significantly accelarated for oniline application.

Compared with previous studies, the unique contributions of this article are summarized as follows.
\begin{enumerate}
  \item A two-stage deep reinforcement learning method is proposed to improve the online safety and efficiency via offline pre-training. The proposed data-driven VVC simultaneously works in model-free manner online and incorporates the knowledge of approximate models offline. This feature makes the proposed method more practical and effective for the real-world applications.
  \item In the offline stage, AMDP is formulated and an adversary agent is trained to exploit the modeling errors of ADN. The introduction of adversary agent makes the trained OFF-A robust to the model mismatch. Instead of conducting the adversarial learning separately as the existing works, this paper proposes a novel JASAC to exploit the shared information between agents, which endows the algorithm with remarkable higher efficiency and convergence. The proposed JASAC algorithm can also be further applied to other DRL-based application.
  \item In the online stage, the OFF-A is transferred to ON-A for continuous learning and application. Since the the OFF-A is robust to the transfer gap due to the adversarial learning offline, the proposed algorithm gains significantly better performance and safety in the online application than the legacy DRL-based VVC.
\end{enumerate}

The remainder of this article is organized as follows. \Cref{sec:prelim} formulates the online and offline stage of the VVC problem for ADNs as a MDP and a AMDP respectively. Then, the detailed introduction to the proposed VVC algorithm and JASAC are presented in \Cref{sec:methods}. In \Cref{sec:numerical} the results of our numerical study are shown and analyzed. Finally, \Cref{sec:conclusion} concludes this article.

\section{Preliminaries}
In this section, we cover the preliminaries of MDP and AMDP, and then formulate the VVC problem in ADNs into MDP and AMDP.
\label{sec:prelim}
\subsection{Markov Decision Process and Reinforcement Learning}
\label{sub:mdp}
In the online stage, the RL agent ON-A learns through the interaction with an environment $\mathcal{E}$.
As a formalization, MDP defined by the tuple $(\mathcal{S},\mathcal{A},p,\mathcal{R},\gamma)$ is utilized to describe the interaction process and suppose the state of the next transition depends only on current state and the action of the agent.
It should be noted that the exact state space $\mathcal{S}$ is often impossible to get in the real world, so we define $\mathcal{O}$ as the observation space for $\mathcal{S}$.
Due to standard conventions for notation, we put $s\in\mathcal{S}$ in places instead of $o\in\mathcal{O}$ and treat the unobserved states as noises.

In this paper, the state space $\mathcal{S}$ and action space $\mathcal{A}$ are continuous, and the unknown state transition probability $\rho: \mathcal{S}\times\mathcal{S}\times\mathcal{A}\rightarrow [0,\infty)$ is the probability density of the next state $s_{t+1}\in\mathcal{S}$ with the current state $s_t\in\mathcal{S}$ and the action $a_t\in\mathcal{A}$, which means $s_{t+1}\sim \rho(\cdot\left|s_t,a_t\right.)$.
The reward function $\mathcal{R}:\mathcal{S}\times\mathcal{A}\rightarrow \mathbb{R}$ quantifies the agent's performance in each transition. We use $r_t=\mathcal{R}(s_t,a_t)$ to denote the reward on each transition. $\gamma\in[0,1]$ is the discount factor trading off current and future rewards. Also, $s_0$ denotes the initial state. 

The goal is to learn a policy that maximizes the total expected discounted rewards, i.e., $\pi^*(a_t|s_t) = \arg\max_{\pi}J(\pi)$ and $J(\pi)=\mathbb{E}\left[\sum_{t=0}^T \gamma^t r_t\right]$. The policy $\pi$ is the stochastic distribution of the action $a_t$ token by the agent under the state $s_t$, i.e., $a_t\sim\pi(\cdot\left|s_t\right.)$.

In order to implement RL algorithms, two value functions $V^\pi(s)$ and $Q^\pi(s,a)$ are defined. $V^{\pi}(s) = \mathop{\mathbb{E}}\limits_{\tau \sim \pi}\left[\sum_{t=0}^T \gamma^t r_t \left| s_0 = s\right. \right]$ is the state-value function representing the expected discounted reward after state $s$ with the policy $\pi$. $Q^{\pi}(s,a) =  \mathop{\mathbb{E}}\limits_{\tau \sim \pi}\left[\sum_{t=0}^T \gamma^t r_t \left| s_0 = s, a_0 = a\right. \right]$ is the action-value function representing the expected discounted reward after taking action $a$ at state $s$ with the policy $\pi$. Here, $\tau \sim \pi$ is the trajectory when applying $\pi$. From the definition, it is obvious that $V^{\pi}(s) = \mathop{\mathbb{E}}\limits_{a\sim \pi}Q^\pi(s,a)$.

\subsection{Adversarial Markov Decision Process}
\label{sub:amdp}
In the offline stage, we train the OFF-A and set up an adversary agent to mock the transfer gap from OFF-A to ON-A. This adversarial setting can be described as a two player discounted zero-sum Markov game\cite{LITTMAN1994157}, in which the protagonist and the adversary (opponent) are involved. In this paper, we formalize this game as an adversarial MDP, which is an expansion of the MDP in \Cref{sub:mdp}. Note $(\cdot)_{p}$ or $(\cdot)^{p}$ to be the variables for the protagonist and $(\cdot)_{o}$ or $(\cdot)^{o}$ for the adversary.

Hence, AMDP can be expressed as the tuple $(\mathcal{S},\mathcal{A}_{p},\mathcal{A}_{o},\rho,\mathcal{R},\gamma)$.
$\mathcal{A}_{p}$ is the action space of the protagonist representing the OFF-A, so $\mathcal{A}_{p}$ equals $\mathcal{A}$ in the MDP. $\mathcal{A}_{o}$ is the action space of the adversary, which mocks the transfer gap such as parameters errors. $\mathcal{S}$ and $\gamma$ remain the same as MDP.

The definition of the other symbols in \Cref{sub:mdp} are expanded respectively: $\rho: \mathcal{S}\times\mathcal{A}_{p}\times\mathcal{A}_{o}\rightarrow [0,\infty)$ is the state transition probability. At some time step $t$, the protagonist picks an action $a_t^p\sim\pi_p(\cdot|s_t)$ while the adversary chooses $a_t^o\sim\pi_o(\cdot|s_t)$. Then the environment transitions to the next state $s_{t+1}\sim\rho(\cdot|s_t,a_t^p,a_t^o)$.

The reward function $\mathcal{R}$ is expanded to $\mathcal{S}\times\mathcal{A}_{p}\times\mathcal{A}_{o}\rightarrow \mathbb{R}$ and $r_t=\mathcal{R}(s_t,a_t^p, a_t^o)$. In AMDP, the protagonist maximizes $J(\pi_p,\pi_o)$ with $\pi_p$ while the adversary minimizes it using $\pi_o$ as shown in \Cref{eq:pippio}.
\begin{equation}
  \label{eq:pippio}
  \begin{split}
    \pi_p^*(a_t^p|s_t) = \arg\max_{\pi_p}\min_{\pi_o}J(\pi_p,\pi_o)\\
    \pi_o^*(a_t^o|s_t) = \arg\min_{\pi_o}\max_{\pi_p}J(\pi_p,\pi_o)
  \end{split}
\end{equation}

\subsection{VVC Problem Formulation}
The VVC problem of ADNs is formulated as MDP for the online stage and ADMP for the offline stage. The detailed VVC problem settings are given in the supplemental file \cite{zSupplemental}. The specific definitions of state space, action space and reward function are designed for both MDP and AMDP as follows.

\subsubsection{State Space}
The actual state space of the ADN is complex and impossible to fully perceive. Hence, in the MDP and AMDP, the state space $\mathcal{S}$ is supposed to be the same as observation space $\mathcal{O}$. $s\in\mathcal{S}$ is defined as a vector $\mathbf{s}=(\mathbf{P},\mathbf{Q},\mathbf{V},t)$. Here $\mathbf{P},\mathbf{Q}$ is the vector of nodal active/reactive power injections $P_i, Q_i(\forall i \in \mathcal{N})$, $\mathbf{V}$ is the vector of voltage magnitudes $V_i(\forall i \in\mathcal{N})$. $t$ is the time step in each episode. 

\subsubsection{Action Space}
For the protagonist OFF-A in the offline stage and ON-A in the online stage, the action space $\mathcal{A} = \mathcal{A}_p$ includes the controllable range of the reactive power outputs of all IB-ERs and SVCs. Such devices provide high speed reactive power support and have large capacity in high penetration areas. Though other slow devices such as shunt capacitors do exist in ADNs, their operation schemes are usually scheduled offline because of their limited allowable daily switching times.

Hence, $\mathbf{a}_p=\mathbf{a}\in\mathcal{A}$ is defined as $(\mathbf{Q}_G, \mathbf{Q}_C)$. $\mathbf{Q}_G$ and $\mathbf{Q}_C$ are the vectors of reactive power generation of IB-ERs and SVCs $Q_{Gi}, Q_{Ci}$ respectively with range $|Q_{Gi}| \leq \sqrt{S_{Gi}^2-\overline{P_{Gi}}^2}$ and $\underline{Q_{Ci}} \leq Q_{Ci} \leq \overline{Q_{Ci}}$.

The action space of adversary in the offline stage is defined as the modeling errors represented by line parameter deviations. For $a_o\in\mathcal{A}_o$, $a_o = (\mathbf{\Delta r}_{ij}, \mathbf{\Delta x}_{ij}), (i,j)\in\mathcal{E}$ where $\mathcal{E}$ is the collection of branches. The simulation parameters are calculated with \Cref{eq:deltarx},
\begin{equation}
  \label{eq:deltarx}
  \begin{split}
    r_{ij} = r^0_{ij} + \Delta r_{ij}, \quad \forall (i,j)\in\mathcal{E} \\
    x_{ij} = x^0_{ij} + \Delta x_{ij}, \quad \forall (i,j)\in\mathcal{E} \\
  \end{split}
\end{equation}
where $r_{ij}, x_{ij}$ are resistance and reactance of branch $ij$, $r^0_{ij}$ and $x^0_{ij}$ are the approximate parameters.

The variation range of $\mathbf{\Delta r}_{ij}$ and $\mathbf{\Delta x}_{ij}$ determines the adversarial force in AMDP.
Unreasonably strong adversary leads to very conservative OFF-A and make the offline adversarial training unstable. 
In ADN, the line parameter is calculated according to the type of wire and its length. Though such parameters are usually influenced by temperature,shape and manufacturing tolerance, their deviations are limited in reality.

\subsubsection{Reward Function}
Though the reward is generated by the environment without $s$ involved explicitly in the classic RL algorithms, it is designed to be a function of previous states (observations). The reward function depends on the goal of VVC and is the key to achieve proper performance.
In this paper, the objectives are minimization of active power loss and mitigation of voltage violations. Hence, the reward consists of two terms: penalty for active power loss $R_P$, and penalty for voltage violations $R_V$.

$R_P$ is calculated according to the active power injections $\mathbf{P}(t)$ as shown in \Cref{eq:rp}:
\begin{equation}
  \label{eq:rp}
  R_P(t) = -\sum_{i\in\mathcal{N}}P_{i}(t)
\end{equation}

In this paper, we focus the grid with high penetration of DGs, where the voltage violations are usually severe and the regulation capacity may be not enough to eliminate all violations in some scenarios.
Hence, $R_V$ is assigned according to the 2-norm of voltage magnitude violations called voltage violation rate (VR) as $R_V(t)=\text{VR}(t)$.
$\text{ReLU}$ is the well-known rectified linear unit function defined as $\text{ReLU}=\max(0, x)$.
We have $R_V(t) \geq 0$ where the equality holds if and only if all voltage magnitudes satisfy the voltage constraints.
Note that $R_V$ can effectively mitigate the voltage violations even when some violating nodes cannot be eliminated.

\begin{equation}
  \begin{split}
    R_V(t) = -\sum_{i\in\mathcal{N}}\left[\text{ReLU}^2(V_i(t)-\overline{V}) + \text{ReLU}^2(\underline{V}-V_i(t))\right]
  \end{split}
\end{equation}

The overall reward is the weighted sum of previous penalty terms, with a hyperparameter $C_v > 0$ as the weight ratio.
\begin{equation}
  r_t = R(s_t,a_t,s_{t+1}) \doteq R_P(t) + C_v R_V(t)
\end{equation}

\section{Methods}
\label{sec:methods}
\subsection{Overview of the Two-stage DRL-based VVC}
\label{sub:overall}
\begin{figure}[!htbp]
  \centering
  \includegraphics[width=\linewidth]{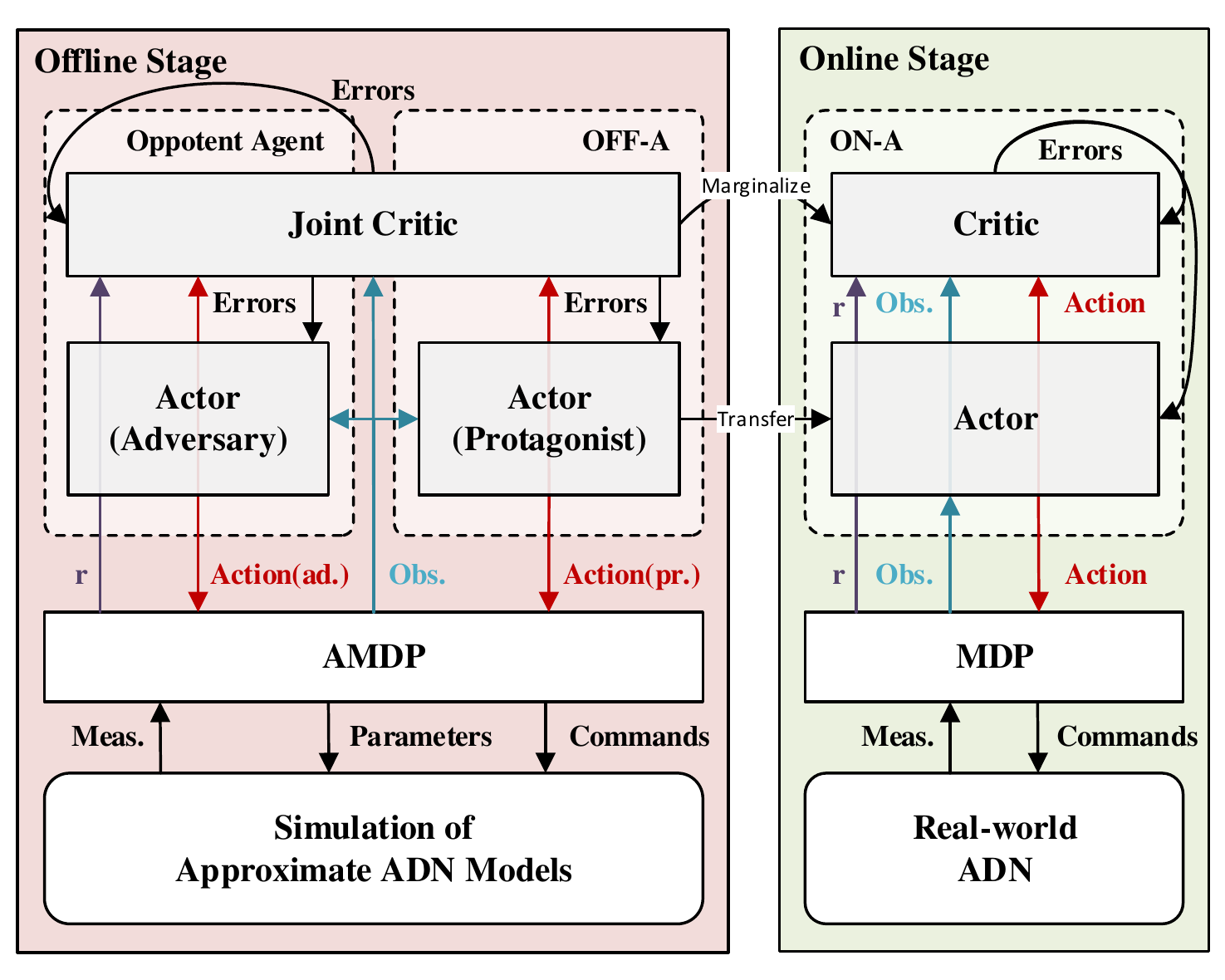}
  \caption{The overall structure of the proposed two-stage VVC algorithm}
  \label{fig:overall}
\end{figure}

In this paper, we propose a two-stage DRL-based VVC with a transferable DRL agent as shown in \Cref{fig:overall}. In the online stage, the state-of-art deep reinforcement learning algorithm soft actor-critic (SAC) is adopted for the ON-A due to its outstanding sample efficiency.
However, the direct usage of SAC for online training often leads to severe security problems or unbearable cost in the start-up phase. Hence, we propose an alternative solution to train the transferable OFF-A in the offline stage which fully utilize the domain knowledge and historical experiences, and transfer OFF-A to ON-A to cope with the online training challenge.

Based on the off-policy feature of SAC that allows the target and behavior policy to be different, it is possible to use two common sources of samples: 1) offline models of ADNs, and 2) historical control samples. Both sources can only provide a rough description of the actual physical system since errors exist in the parameters of models and current system varies from the historical one over time. In this paper, we focus on the former source since the latter one is not always available.

In this paper, the OFF-A is trained against the possible modeling errors to be robust to the transfer gap or transferable in another way. The space of all possible modeling errors is termed the disturbance space $\mathcal{X}$. Instead of requiring the exact distribution of $\mathcal{X}$ and sampling all possible combinations \cite{rajeswaran2016epopt}, an adversary agent is trained jointly to mock the modeling errors and impede the protagonist by applying disturbances on the model parameters. The aim of the adversary is to disturb the protagonist in certain $\mathcal{X}$ and minimize the total discounted reward that the protagonist can get. Therefore, the adversary agent can learn the most tough scenarios for the protagonist and make the latter one as robust as possible regarding $\mathcal{X}$.

Though the original adversarial RL method \cite{pinto2017robust} has been justified valid, it suffers poor efficiency and convergence in large scale cases. To improve the performance, we propose an adversarial reinforcement learning algorithm JASAC in \Cref{sub:JASAC}. Instead of training the protagonist and the adversary separately, JASAC shares the estimation of the value functions between the two agents, since they are in the same zero-sum game. The shared information allows each agent to optimize the value function approximation considering the other's behavior, so both the efficiency and convergence are promoted especially in cases with large scale of action and state space.

Finally, after the offline stage, the robust OFF-A is transferred to ON-A in the online stage. The numerical tests show that ON-A can realize VVC of the ADN safely and efficiently with continuous learning from online samples.

\subsection{Soft Actor-Critic}
In order to solve the MDPs, the actor-critic framework is popular among modern RL algorithms.
Well-known actor-critic methods includes proximal policy optimization (PPO)\cite{schulman2017proximal}, asynchronous advantage actor-critic (A3C)\cite{mnih2016asynchronous}, deep deterministic policy gradient (DDPG)\cite{silver2014deterministic} and SAC\cite{haarnoja2018sac,haarnoja2018learning}.
PPO and A3C are on-policy methods, which means the new samples have to be generated according to the latest policy of the agent.
The sample efficiency of such methods is often far from satisfactory.
Though DDPG is a famous off-policy method, it learns a deterministic policy and suffers in the online exploration.

To cope with these challenges, SAC follows the previous maximum entropy algorithms especially soft Q-learning (SQL) \cite{haarnoja2017reinforcement} and provides an off-policy maximum entropy RL algorithm. The state value function is entropy-regularized as
\begin{equation}
    V^{\pi}(s) = \mathop{\mathbb{E}}\limits_{\tau \sim \pi}\left[ \left. \sum_{t=0}^{\infty} \gamma^t \bigg( r_t + \alpha H\left(\pi(\cdot|s_t)\right) \bigg) \right| s_0 = s\right]
\end{equation}
where $H(\pi(\cdot|s_t))=\mathop{\mathbb{E}}\limits_{a\sim\pi(\cdot|s_t)}\left[-\log \pi(\cdot|s_t)\right]$ is the entropy for the stochastic policy at $s_t$.
And $Q^{\pi}(s,a)$ is entropy-regularized accordingly as $Q^{\pi}(s,a) = \mathop{\mathbb{E}}\limits_{\tau \sim \pi} \bigg[\sum_{t=0}^T \gamma^t r_t + \alpha\sum_{t=1}^T \gamma^t H\big(\pi(\cdot|s_t)\big) \big| s_0 = s, a_0 = a \bigg]$.

During the learning process, all the samples of MDP are stored in the replay buffer $\mathcal{D}$ as $(s,a,r,s')\in \mathcal{D}$. To approximate $Q^{\pi}$ by neural network iteratively, Bellman equation is applied to the entropy-regularized $Q^{\pi}$ and approximated with samples as \Cref{eq:bellman} since $\mathbb{E}_{s', a'}H\left(\pi\left(\cdot|s'\right)\right)=-\mathbb{E}_{s', a'}\log \pi\left(\cdot|s'\right)$. 
\begin{equation}
  \label{eq:bellman}
  \begin{aligned}
    Q^{\pi}(s,a) &=
    \mathop{\mathbb{E}}\limits_{s', a'}
    \Big[R(s,a,s') + \gamma \left(Q^{\pi}\left(s',a'\right) + \alpha H\left(\pi\left(\cdot|s'\right)\right)\right) \Big] \\
    & \approx r + \gamma \left( Q^{\pi}\left(s', \tilde{a}'\right) - \alpha \log \pi \left(\left. \tilde{a}'\right|s' \right)  \right), \tilde{a}'\sim \pi(\left.\cdot \right| s')
  \end{aligned}
\end{equation}
Note that $\tilde{a}'$ is generated from the latest $\pi(\left.\cdot\right|s')$ while $s'$ is from $\mathcal{D}$. With this approximation, we use mean-squared Bellman error (MSBE) function to update the $Q$ network.

With the value function as the ``critic'', the ``actor'' policy is optimized to maximize the state value function $V^\pi(s)$. In SAC, the reparameterization trick using a squashed Gaussian policy is introduced: $\tilde{a}_{\theta}(s, \xi) = \tanh\left( \mu_{\theta}(s) + \sigma_{\theta}(s) \odot \xi \right), \, \xi \sim \mathcal{N}(0, \mathbf{I})$, where $\mu_{\theta},\sigma_{\theta}$ are two parameterized neural networks. Hence, the expectation over actions is converted to the expectation over noise $\xi$, and the policy parameter $\theta$ can be optimized according to
\begin{equation}
  \label{eq:sacpi}
  \max_{\theta} \mathop{\mathbb{E}}\limits_{\substack{s\sim \mathcal{D}\\ \xi\sim \mathcal{N}}} \left[ Q\left(s,\tilde{a}_{\theta}(s,\xi)\right) - \alpha \log \pi_{\theta}\left(\left.\tilde{a}_{\theta}(s,\xi)\right|s\right) \right]
\end{equation}

A naive implement of SAC is given in the supplemental file \cite{zSupplemental} using \Cref{eq:bellman,eq:sacpi}. Several techniques \cite{haarnoja2018learning} to make SAC more practical and stable, such as optimization of the $\alpha$, frozen target and two Q-functions, are omitted here.

\subsection{Jointly Adversarial Soft Actor-Critic}
\label{sub:JASAC}
In the offline stage, the protagonist and adversary agents are trained respectively to solve the AMDP in \Cref{sub:amdp}. To realize the adversarial training, the original algorithm robust adversarial reinforcement learning (RARL) \cite{pinto2017robust} alternatively trains the agents in a separate manner using policy gradient method. Though the effectiveness of RARL has been tested under several standard robotic environments, its stability, convergence and efficiency remain concerns. The bottlenecks for RARL are: 1) the separate training structure blocks the knowledge sharing between agents and make the algorithm less efficient; 2) policy gradient method is on-policy and suffers from stability and sample efficiency.

In this work, we innovate an algorithm called jointly adversarial soft actor-critic (JASAC) for the AMDP. The overall framework is provided in \Cref{alg:jasac}. An off-policy maximum entropy optimization is conducted based on SAC for each agent in an alternative manner. During the training process, the agents share the knowledge of the system using a spanned state-action value function and each optimizes the policy considering the other one on each step, which is the key novelty of JASAC.

Under the formulation of AMDP, we incorporate the maximum entropy framework like SAC for both agents. The value function is derived as
\begin{equation}
  \label{eq:ad_v}
  \begin{split}
    &V^{\pi}(s) = \mathbb{E}\Big[
    \sum_{t=1}^{T} \gamma^{t-1}
    \Big(
    \mathcal{R}(s_t,a_t^{p},a_t^{o})\\
    &-\alpha_{p}\log\pi_{p}(a^{p}_t|s_t)
    -\alpha_{o}\log\pi_{o}(a^{o}_t|s_t)
    \Big)
    \Big]
  \end{split}
\end{equation}
where $\pi$ denotes $(\pi_p,\pi_o)$, and $\alpha_{p},\alpha_{o}$ are the Lagrange multipliers for the entropy constraints. $\alpha_{p} > 0$ and $\alpha_{o} < 0$ since the adversary minimizes the expectation of total discounted reward as \Cref{eq:pippio}.

\subsubsection{Learning Q-Functions}
In a separate training framework, each agent implements an approximator $Q_\phi(s,a)$ for the state-action value function. Though the action space of the agents are not the same, they are playing in the same game so that the actual value function $V^*(s)$ should be identical as $\max_{\pi_p} \min_{\pi_o} V^{\pi_{p}\pi_{o}}(s) = \min_{\pi_o} \max_{\pi_p} V^{\pi_{p}\pi_{o}}(s)$ \cite{grau2018balancing}. The consistent value function inspires us to share the value approximation between agents during adversarial training and achieve the equilibrium faster. To do this, we define a joint Q-function instead of two separate ones as $Q^\pi:\mathcal{S}\times\mathcal{A}_p\times\mathcal{A}_o\rightarrow \mathbb{R}$ according to \Cref{eq:ad_v}. 

Typically, we adapt Bellman equation to estimation the Q-value as shown in \Cref{eq:qbell2}. The shared information in $Q^\pi$ allows us to consider both current policies $\pi_p,\pi_o$ during the approximation instead of doing it alternatively.
\begin{equation}
  \label{eq:qbell2}
  \begin{split}
    Q^{\pi}(s,a_p,a_o)
    =& \mathop{\mathbb{E}}\limits_{s',a'_p,a'_o}\Big[R(s,a_p,a_o,s') + \gamma\big(Q^{\pi}(s',a'_p, a'_o)\\
     &- \alpha_p \log \pi_p(a'_p|s') - \alpha_o \log \pi_o(a'_o|s') \big)\Big]
  \end{split}
\end{equation}

Also, the parameters of target Q networks $\hat\phi$ are delayed comparing to the learned $\phi$ as indicated in \cite{haarnoja2018sac}.
Hence, the target of Q-functions $y(r, s')$ is calculated using the Bellman equation \Cref{eq:qbell2} as
\begin{equation}
  \begin{split}
    y(r, s') = r + \gamma\Big[ & Q_{\hat\phi}(s',\tilde{a}'_p,\tilde{a}'_o) \\
      & - \alpha_p \log \pi(\tilde{a}'_p|s')
       - \alpha_o \log \pi(\tilde{a}'_o|s') \Big]
  \end{split}
\end{equation}
where $\tilde{a}'_p$ and $\tilde{a}'_o)$ are generated using the latest $\pi_p(\cdot|s')$ and $\pi_o(\cdot|s')$. With this target value, we update $\phi_1,\phi_2$ by gradient descent on MSBE function $J_Q(\phi)$:
\begin{equation}
  J_Q(\phi) = \mathop{\mathbb{E}}\limits_{(s,a,r,s') \sim {\mathcal D}} \left[
    \big( Q_{\phi}(s,a) - y(r,s') \big)^2
    \right]
\end{equation}

\subsubsection{Learning the Policy}
The policies are learnt by optimizing the expected value. With the joint Q-functions, the value function $V^{\pi}(s)$ is calculated by the expectation on both action spaces. 
\begin{equation}
  \label{eq:vpi2}
  \begin{split}
    V^{\pi}(s) = \mathop{\mathbb{E}}\limits_{\substack{a_p\sim\pi_p \\ a_o\sim\pi_o}}
    \Big[&Q^{\pi}(s,a_p,a_o) \\
    &- \alpha_p \log \pi_p(a_p|s)  - \alpha_o \log \pi_o(a_o|s)\Big]
  \end{split}
\end{equation}

To decouple the expectation and actions, the reparameterization trick in SAC is utilized here as $\tilde{a}_{p}^\theta(s, \xi_p) = \tanh\left( \mu_{\theta}(s) + \sigma_{\theta}(s) \odot \xi_p \right), \xi_p \sim \mathcal{N}(0, \mathbf{I})$ and $\tilde{a}_{o}^\omega(s, \xi_o) = \tanh\left( \mu_{\omega}(s) + \sigma_{\omega}(s) \odot \xi_o \right), \xi_o \sim \mathcal{N}(0, \mathbf{I})$, where $\theta$ is the parameter of protagonist policy and $\omega$ of adversary policy. The original problem $\max_{\pi_p}\min_{\pi_o}V^\pi(s_0)$ is transformed to
\begin{equation}
  \begin{split}
    &\max_{\theta} \min_\omega \mathop{\mathbb{E}}\limits_{\substack{s \sim \mathcal{D} \\ \xi_p \sim \mathcal{N} \\ \xi_o \sim \mathcal{N}}}
    \Big[Q_\phi \left( s,\tilde{a}^{\theta}_p(s,\xi_p),\tilde{a}^{\omega}_o(s,\xi_o) \right) \\
    & - \alpha_p \log \pi_p(\tilde{a}^{\theta}_p(s,\xi_p)|s)  - \alpha_o \log \pi_o(\tilde{a}^{\omega}_o(s,\xi_o)|s)\Big]
  \end{split}
\end{equation}

Though it is possible in SAC to determine the optimal policy directly if the action space is discrete and small, evaluating the equilibrium solution in every step is time-consuming for a minimax optimization problem on continuous space. Following SAC and RARL, we alternatively learn $\pi_p$ and $\pi_o$ in a descent manner. Because $\theta$ and $\omega$ are independent in the entropy terms, we derive the loss function for policy parameters $\theta,\omega$ as
\begin{equation}
  \label{eq:jpip}
  \begin{split}
    J_{\pi_p}(\theta) = \frac{1}{|B|}\sum_{s\in B}\Big[&
    Q_{\hat\phi}(s,\tilde{a}^{\theta}_p(s,\xi_p),\tilde{a}^{\omega}_o(s,\xi_o)) \\
    & - \alpha_p \log \pi_p(\tilde{a}^{\theta}_p(s,\xi_p)|s)\Big]
  \end{split}
\end{equation}
\begin{equation}
  \label{eq:jpio}
  \begin{split}
    J_{\pi_o}(\omega) = \frac{1}{|B|}\sum_{s\in B}\Big[&
    Q_{\hat\phi}(s,\tilde{a}^{\theta}_p(s,\xi_o),\tilde{a}^{\omega}_o(s,\xi_o)) \\
    & - \alpha_o \log \pi_o(\tilde{a}^{\omega}_o(s,\xi_o)|s)\Big]
  \end{split}
\end{equation}
where $B$ is a batch of samples from $\mathcal{D}$. It should be emphasised that in \Cref{eq:jpip,eq:jpio}, the joint Q-functions provides the possible state value considering not only one's own action, but also the opponent's action. It makes the updated policy take the other's stochastic actions into consideration and correct the search direction.

Finally, before migrating the protagonist to the online stage, we have to bridge JASAC and SAC. The protagonist policy $\pi_p$ can be migrated directly as $\pi$ in SAC. Because we want to continue the training online to realize a model-free VVC, $Q^*_{\phi}$ is marginalized for protagonist as $Q^*_{\Phi}$ since no adversary is conducted online.
\begin{equation}
  \label{eq:margin}
  Q^*_{\Phi}(s, a_p) = \mathop{\mathbb{E}}\limits_{\xi_o\sim\mathcal{N}}Q^*_{\phi} (s, a_p, \tilde{a}^{\omega}_o(s,\xi_o))
\end{equation}

\begin{algorithm}
  \caption{Jointly Adversarial Soft Actor-Critic}
  \label{alg:jasac}
  Initialize experience pool, function approximators' parameter vectors $\theta, \omega$ for policy and $\phi$ for value function\;
  \ForEach{
    episode
  }{
    \ForEach{
      environment step
    }{
      $a_t^p\sim\pi_p(\cdot|s_t), a_t^o\sim\pi_o(\cdot|s_t)$\;
      Feed $a_t^p,a_t^o$ to the environment, get reward $r_t$ and next state $s_{t+1}$\;
      $\mathcal{D}\leftarrow\mathcal{D}\cup\{(s_t,(a_t^p,a_t^o),r_t,s_{t+1})\}$\;
    }
    \ForEach{gradient step}{
      $\phi \leftarrow \phi - \lambda \nabla_{\phi}J_Q(\phi)$ \;
      $\theta \leftarrow \theta + \lambda \nabla_{\theta}J_{\pi_p}(\theta)$\;
      $\omega \leftarrow \omega - \lambda \nabla_{\omega}J_{\pi_o}(\omega)$\;
    }
    $\hat\phi \leftarrow \eta \phi + (1-\eta) \hat\phi$\;
  }
  Marginalize $Q^*_{\phi}$ for protagonist as $Q^*_{\Phi}$ using \Cref{eq:margin}\;
  \KwOut{Transferable protagonist state-action value function $Q^*_{\Phi}$ and $\pi^*_{p}$ for online stage.}
\end{algorithm}

\section{Numerical Study}
\label{sec:numerical}
In this section, numerical experiments are conducted using IEEE 33-bus and 69-bus distribution test cases to validate the advantage of the proposed two-stage method over some popular benchmark algorithms including DRL algorithms and optimization-based algorithms. Steady-state distribution system reinforcement learning environments are built under the scheme of the well-known toolkit Gym\cite{brockman2016openai}. Detailed simulation configuration is given in the supplemental file \cite{zSupplemental}.

\subsection{Proposed and Baseline Algorithms Setup}
In the offline stage, the proposed algorithm JASAC is implemented for solving AMDP. For the benchmark algorithm, we first follow the structure of RARL. The original version of RARL utilized on-policy algorithms to train both protagonist and adversarial agents and showed poor convergence in our environments. Hence, we substitute the on-policy algorithm for both agents with off-policy algorithms, which can be seen as a separately adversarial soft actor-critic (ASAC) algorithm. The major difference between JASAC and ASAC is that the former one shares critic value information between agents. An optimization-based algorithm with SOCP relaxation is implemented using the power flow model \cite{zSupplemental}, either with oracle parameters (VVO) or with fake parameters (VVO with error). Also, in order to show the effectiveness of adversarial training, SAC is selected as one of the benchmarks to solve MDP.

In the online stage, in addition to optimization-based and SAC benchmarks, we also migrate the trained SAC in the offline stage to the online (preSAC) following our two-stage DRL structure. The algorithm hyperparameters for both stages are listed in the supplemental file \cite{zSupplemental}.

Due to the stochastic property of DRL-based algorithms, we use 5 independent random seeds for each group of experiment, whose mean value and error bounds are presented in the figures.

\subsection{Offline Convergence and Efficiency}
\begin{figure}[htbp]
  \centering
  \subfloat[IEEE 33-bus system\label{fig:33offJoint}]{\includegraphics[width=0.5\linewidth]{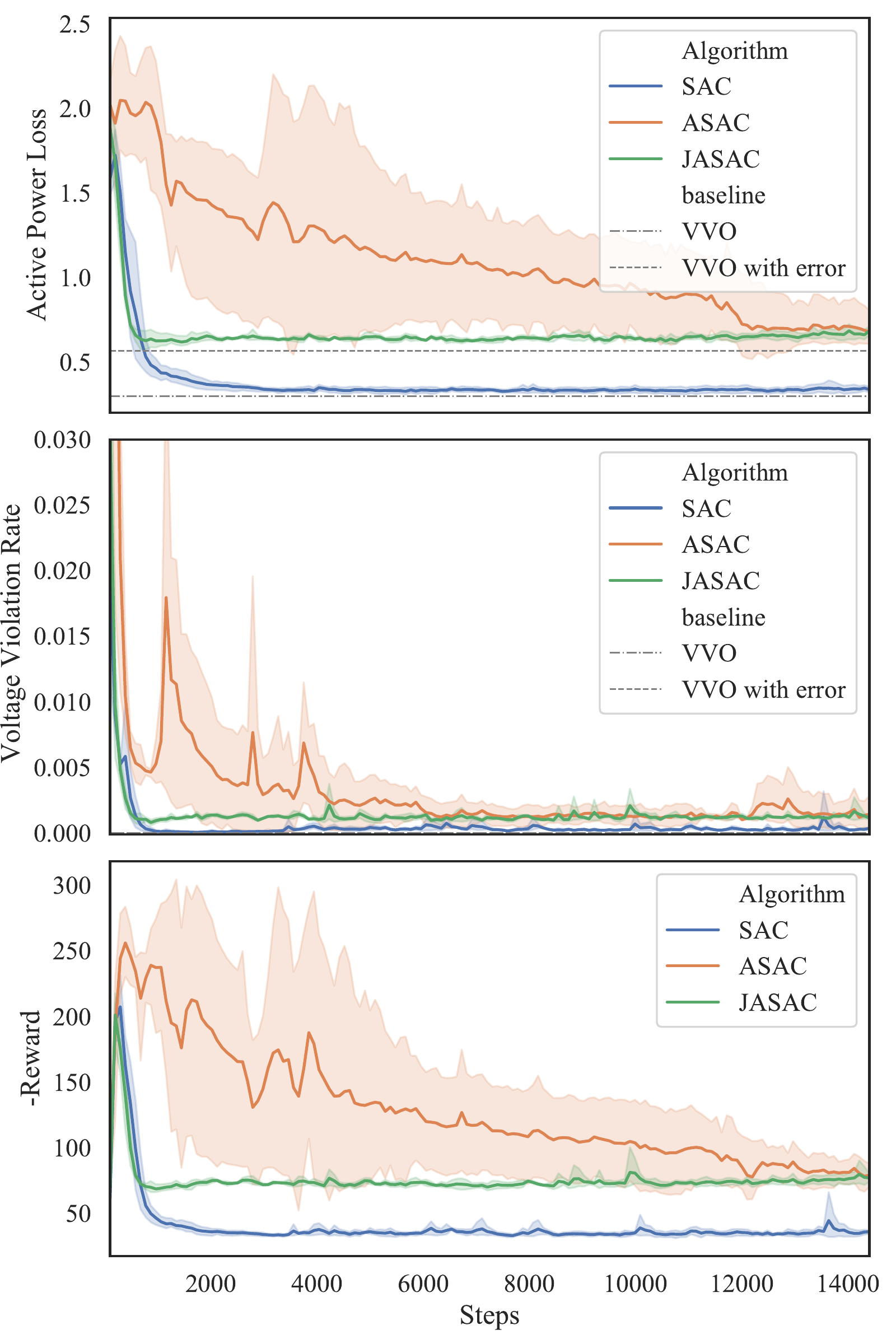}}
    \hfill
  \subfloat[IEEE 69-bus system\label{fig:69offJoint}]{\includegraphics[width=0.5\linewidth]{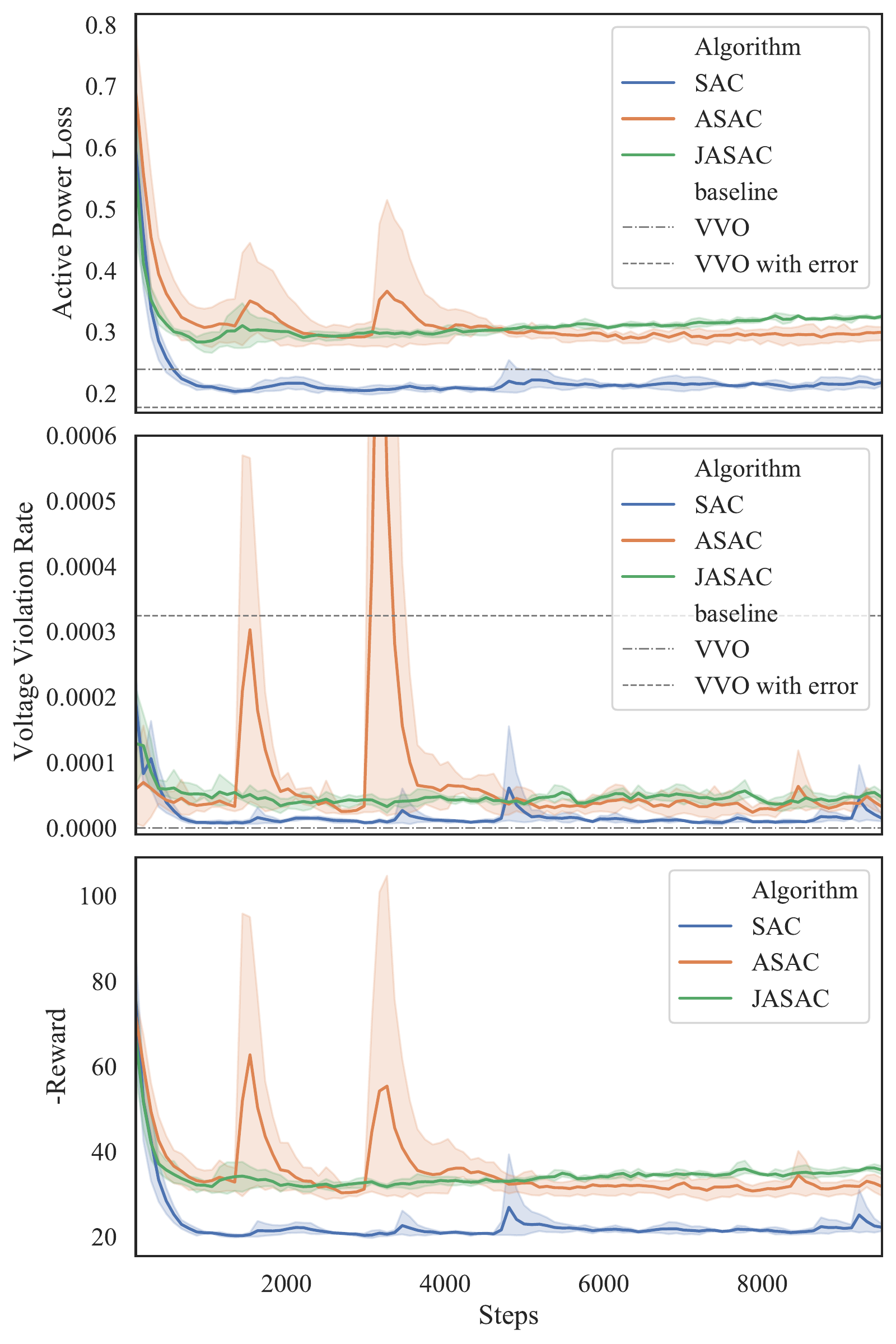}}
  \caption{Offline training process with the test systems}
  \label{fig:offJoint} 
\end{figure}

In the offline stage, we train the proposed JASAC algorithm and DRL-based benchmarks with our simulation environments. Also, optimization-based benchmarks are evaluated with corresponding models respectively. The step average value of active power loss, voltage violation rate and negative reward (only for DRL algorithms) are shown in sub-figures of \Cref{fig:offJoint}. The solid curves (green for JASAC, red for ASAC, blue for SAC) represent the mean performance across independent random seeds, and the light color filled regions are corresponding error bounds. The optimization-based benchmarks are drawn as black dashed lines.

The first important observation from \Cref{fig:offJoint} is that SAC algorithm converges to a lower active power loss than the optimization-based method VVO without oracle parameters, which significantly  reveals the advantage of DRL-based algorithms over such parameter-sensitive optimization method regarding Volt/VAR control problem. Such fact could also be supported by the follow-up online results (\Cref{fig:onJoint3,tbl:online33,tbl:online69}) and other related works \cite{8944292,li2019coordination}.
On the other hand,  though the oracle VVO attains the minimum of active power loss theoretically once given all true parameters,  SAC algorithm could closely approach it after certain iterations, as depicted in the figure.

In terms of the comparison between JASAC/ASAC and SAC, we notice that the first two methods converge to the same value above SAC. Such gap arises due to the extra adversarial training mechanism in JASAC/ASAC, in which the adversary seeks for worse cases using model errors, essentially sacrificing a bit training performance for the robustness of the protagonist. In \Cref{sub:online}, we would further discuss that such design would help to produce trust-worth actions at the very start when dealing with the transfer gap from offline to online. Fortunately, such trade-off for robustness is almost of no cost, since in offline stage, our simulation has no actual loss on the real world system.

Finally, comparing JASAC with ASAC, the advantages of JASAC over ASAC with respect to training efficiency and convergence rate are evidently shown in \Cref{fig:offJoint}. The third subplot, which illustrates the change of negative rewards with iterations, well reveals the dynamics of the training process. We could see that even with adversarial training, JASAC achieves similar efficiency and convergence rate as SAC,  while ASAC suffers from the balancing process in the adversarial training. In fact, this significant improvement of JASAC compared to ASAC is credited to the the shared value information of the agents and the consideration of each other's policy. Such features  make  JASAC preferable in practice  for  Volt/VAR  control, not only in this study but also in more complex potential tasks.

\subsection{Online Safety and Efficiency}
\label{sub:online}

\begin{figure}[!htbp]
  \centering
  \subfloat[IEEE 33-bus system\label{fig:33onJoint3}]{\includegraphics[width=0.5\linewidth]{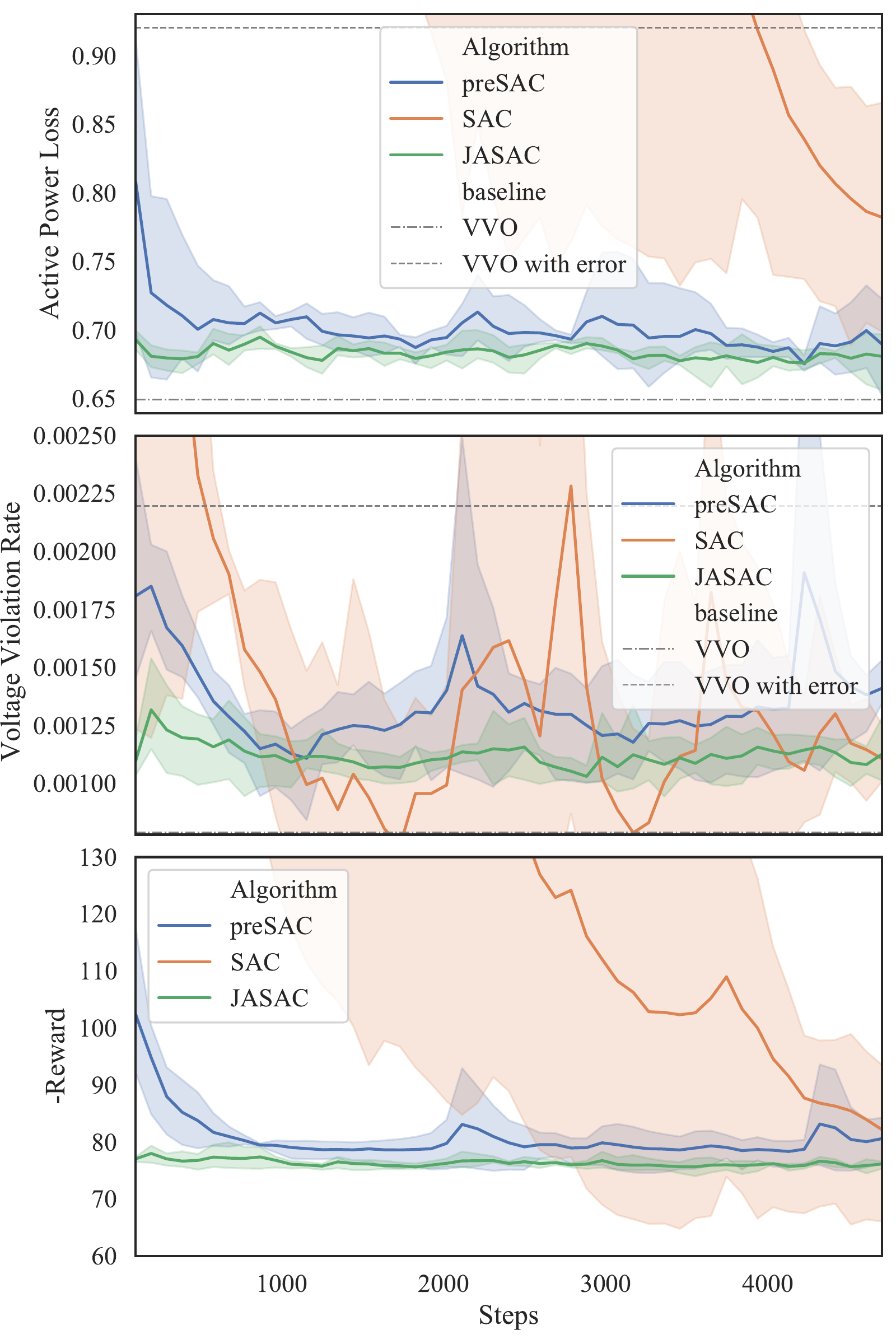}}
    \hfill
  \subfloat[IEEE 69-bus system\label{fig:69onJoint3}]{\includegraphics[width=0.5\linewidth]{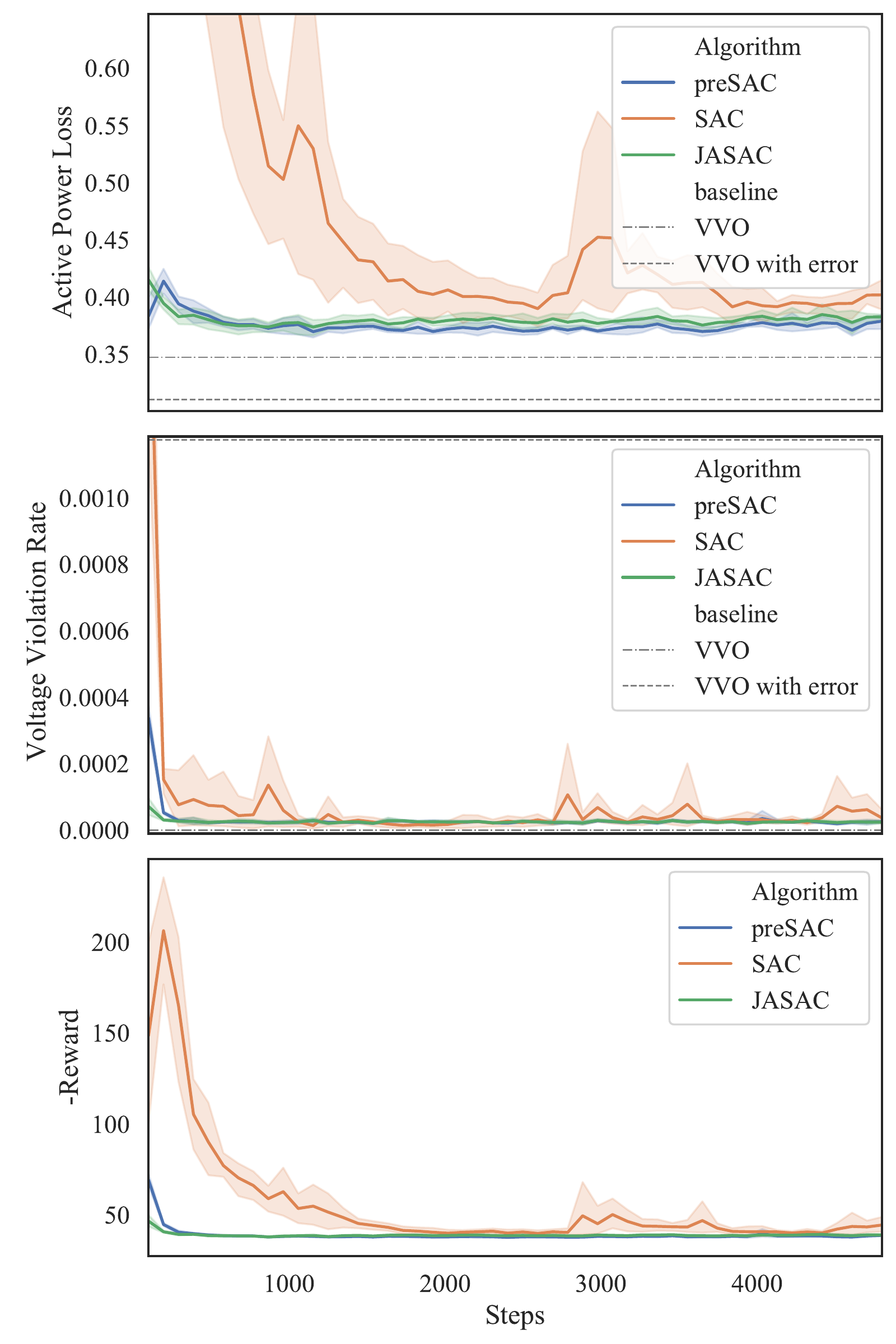}}
  \caption{Online Volt/VAR control process with the test systems}
  \label{fig:onJoint3} 
\end{figure}

In the online stage, while maintaining similar simulation ideas as the offline stage, we change our focus to real world model, where simulations represent real world control process and actions are no longer cost-free. Hence, safety and efficiency become more important in every part of the simulations.  \Cref{fig:onJoint3} shows an example line plot and \Cref{tbl:online33,tbl:online69} includes some major indices.  In the table, we calculate the start and max value of active power loss and VR. The ideal optimal solution calculated by VVO with oracle model and $\text{VR} = 0$ is used as a baseline to calculate the increments in active power loss of other methods. Note that we trained the stochastic algorithms with 5 independent random seeds, based on which we could get the mean values and standard deviations across  seeds.

\begin{table}[!htbp]
  \centering
  \begin{threeparttable}
    \caption{Online performance comparison with the 33-bus system}
    \label{tbl:online33}
    \begin{tabular}{@{}crrrrrr@{}}
      \toprule
      \multirow{2}{*}{} & \multicolumn{1}{c}{\multirow{2}{*}{Algo.}} & \multicolumn{3}{c}{$P_\text{loss}/\text{MW}$}    & \multicolumn{2}{c}{VR}      \\ \cmidrule(l){3-7} 
      & \multicolumn{1}{c}{}    & \multicolumn{1}{c}{Mean} & \multicolumn{1}{c}{Inc.\tnote{*}} & Std. & Mean & Std. \\ \midrule
      \multirow{4}{*}{Start} & SAC       & 2.39e-00 & 1.74e-00 & 2.31e-01 & 3.58e-03 & 3.03e-03 \\
                             & preSAC    & 8.08e-01 & 1.59e-01 & 1.08e-01 & 1.81e-03 & 5.18e-04 \\
                             & JASAC     & 6.93e-01 & 4.37e-02 & 6.94e-03 & 1.10e-03 & 8.41e-05 \\
                             & VVO\tnote{+} & 9.21e-01 & 2.71e-01 & -        & 2.19e-03 & -        \\ \midrule
      \multirow{4}{*}{Max}   & SAC       & 2.41e-00 & 1.76e-00 & 2.05e-01 & 1.08e-02 & 1.22e-02 \\
                             & preSAC    & 8.28e-01 & 1.79e-01 & 7.87e-02 & 2.40e-03 & 6.54e-04 \\
                             & JASAC     & 7.00e-01 & 5.02e-02 & 3.23e-03 & 1.39e-03 & 1.34e-04 \\
                             & VVO\tnote{+} & 9.21e-01 & 2.71e-01 & -        & 2.19e-03 & -        \\ \midrule
      \multicolumn{2}{l}{Optimal Sol.\tnote{o}}    & 6.49e-01 & 0 & - & 0 & - \\ \bottomrule
    \end{tabular}
    \begin{tablenotes}
      \item[*] mean value increment comparing to the lower bound.
      \item[+] optimization-based VVO with the offline approximate model.
      \item[o] ideal optimal solution using perfect ADN model.
    \end{tablenotes}
  \end{threeparttable}
\end{table}

\begin{table}[!htbp]
  \centering
  \begin{threeparttable}
    \caption{Online performance comparison with the 69-bus system}
    \label{tbl:online69}
    \begin{tabular}{@{}crrrrrr@{}}
      \toprule
      \multirow{2}{*}{} & \multicolumn{1}{c}{\multirow{2}{*}{Algo.}} & \multicolumn{3}{c}{$P_\text{loss}/\text{MW}$}    & \multicolumn{2}{c}{VR}      \\ \cmidrule(l){3-7} 
      & \multicolumn{1}{c}{}    & \multicolumn{1}{c}{Mean} & \multicolumn{1}{c}{Inc.\tnote{*}} & Std. & Mean & Std. \\ \midrule
      \multirow{4}{*}{Start} & SAC       & 1.72e-00 & 1.37e-00 & 2.30e-01 & 1.74e-03 & 6.50e-04  \\
                             & preSAC    & 3.84e-01 & 3.60e-02 & 1.89e-02 & 3.38e-04 & 5.02e-05  \\
                             & JASAC     & 4.15e-01 & 6.75e-02 & 1.38e-02 & 7.16e-05 & 3.31e-05  \\
                             & VVO\tnote{+} & 3.11e-01 & -3.7e-02 & - & 1.18e-03 & -  \\ \midrule
      \multirow{4}{*}{Max}   & SAC       & 1.96e-00 & 1.61e-00 & 3.83e-01 & 1.74e-03 & 6.50e-04  \\
                             & preSAC    & 4.15e-01 & 6.71e-02 & 1.28e-02 & 3.38e-04 & 5.02e-05  \\
                             & JASAC     & 4.15e-01 & 6.75e-02 & 1.38e-02 & 7.31e-05 & 3.15e-05  \\
                             & VVO\tnote{+} & 3.11e-01 & -3.7e-02 & - & 1.18e-03 & -  \\ \midrule
      \multicolumn{2}{l}{Optimal Sol.\tnote{o}}    & 3.48e-01 &        0 & -        & 0        & -         \\ \bottomrule
    \end{tabular}
    \begin{tablenotes}
      \item[*] mean value increment comparing to the lower bound.
      \item[+] optimization-based VVO with the offline approximate model.
      \item[o] ideal optimal solution using perfect ADN model. 
    \end{tablenotes}
  \end{threeparttable}
\end{table}

The advantage of the proposed two-stage structure for DRL-based algorithms emerges in the online plot \Cref{fig:onJoint3} and \Cref{tbl:online33,tbl:online69}, where we could clearly observe a much better performance of JASAC and preSAC than SAC. The rationale behind this is that the basic rules of the  control tasks and the neural network structures could be well learned in offline training stage and then inherited by the online stage. These consistent prior knowledge would make the algorithms take good actions even at the very beginning of the control process, thus giving rise to a significant improvement of the efficiency.

Further more,  the adversarial training in the offline stage, as is mentioned previously, proves itself to be effective in the online stage when we look into the performances of preSAC and JASAC, especially at their starting phase. Specifically speaking in \Cref{fig:33onJoint3} and \Cref{tbl:online33}, at the start of 33-bus case, preSAC has an active power loss increment of $1.59\times 10^{-1}$ from the ideal optimal solution, while JASAC cuts it by over $70\%$ to $4.37\times 10^{-2}$. JASAC also reduces the VR by nearly $40\%$ to $1.1\times 10^{-3}$. For the max value in the process and other cases, such comparisons still hold.  The key of our proposed JASAC lies in the strategy to prepare the agent for the transfer gap by first sacrificing a bit training performance in the cost-free offline stage, in return for the robustness of the protagonist who could take preeminent actions in the real world online stage.

\section{Conclusion}
\label{sec:conclusion}
A two-stage deep reinforcement learning algorithm is proposed to optimize the reactive power distribution in inverter-based ADNs without accurate model parameters. The key novelty of the proposed algorithm lies in transfer of knowledge from the offline stage to online and significantly improvement on online control safety and efficiency. In the offline stage, we propose a novel RL algorithm JASAC with the formulation of AMDP to make the offline agent robust to the transfer gap. And in the online stage, OFF-A is transferred as ON-A and performs continuous online learning and control using SAC. Numerical studies on IEEE 33-bus and 69-bus test cases have indicated that the proposed two-stage method outperforms the benchmark methods regarding safety and efficiency in the online stage, and also demonstrated that the proposed JASAC has better convergence and efficiency in the offline stage comparing to the benchmark adversarial learning algorithm.

In the future work, the application of the proposed adversarial JASAC to more complex control problems is also a promising research direction.

\ifCLASSOPTIONcaptionsoff
  \newpage
\fi

\bibliographystyle{IEEEtran}
\bibliography{paper}
\end{document}


\title{Supplemental File for \\ Two-stage Deep Reinforcement Learning for Inverter-based Volt-VAR Control in Active Distribution Networks}
\author{Haotian~Liu, ~\IEEEmembership{Student Member},
  Wenchuan~Wu,~\IEEEmembership{Senior Member,~IEEE}}
\maketitle

\section{VVC Problem in Active Distribution Networks}
\label{sub:vvcinadn}
An ADN with $n+1$ nodes can be depicted by an undirected graph $\Pi(\mathcal{N}, \mathcal{E})$ with the collection of all nodes $\mathcal{N} = {0,...,n}$ and the collection of all branches $\mathcal{E} = {(i,j)\in \mathcal{N}\times\mathcal{N}}$. The point of common coupling (PCC) is located at node 0 with a substation connected to the power grid. Typically, the ADN operates radially as a tree with the root at PCC. Since it is common for the ADN in the real world to equip only with single-phase steady-state measurements, the VVC problem is formulated on balanced radial networks for real-time steady-state dispatch in this paper. Noted that since OFF-A does not require the inner details of the model and utilizes only the input and output data, it can be easily extended to unbalanced multi-phase networks.

The system is equipped with $n_\text{IB}$ IB-ERs and $n_\text{CD}$ compensation devices such as static Var compensators (SVC). Without loss of generality, we assume that the IB-ERs and compensation devices are installed on different nodes in $\mathcal{N}$. Accordingly, the collection of the nodes equipped with IB-ERs and compensation devices are noted as $\mathcal{N}_\text{IB}$ and $\mathcal{N}_\text{CD}$.

In this paper, OFF-A is trained offline using a simulation model with approximate parameters to generate data. While we consider the steady-state voltage control, the power flow equations are employed as shown in \Cref{eq:pf}, where $P_{ij},Q_{ij}$ is the active and reactive power flow from node $i$ to $j$, $V_i$ is the voltage at node $i$ and $G_{ij}+jB_{ij} = 1 / (r_{ij}+ j x_{ij})$ is the admittance of branch ${ij}$, and $G_{sh,i} + j B_{sh, i}$ is the shunt admittance of node $i$.

\begin{equation}
  \label{eq:pf}
  \begin{split}
    P_{ij} &= G_{ij}V_i^2 - G_{ij}V_i V_j \cos{\theta_{ij}} - B_{ij}V_i V_j \sin{\theta_{ij}}, \forall ij \in \mathcal{E} \\
    Q_{ij} &= -B_{ij}V_i^2 + B_{ij}V_i V_j \cos{\theta_{ij}} - G_{ij}V_i V_j \sin{\theta_{ij}}, \forall ij \in \mathcal{E} \\
    \theta_{ij} &= \theta_i-\theta_j,\forall ij \in \mathcal{E}
  \end{split}
\end{equation}

Since $\mathcal{N}_\text{IB}\cap \mathcal{N}_\text{CD} = \emptyset$, the power injections at each nodes can be determined via \Cref{eq:injection}.
\begin{equation}
  \label{eq:injection}
  \begin{aligned}
    G_{sh,i} V_i^2 + \sum_{j \in K(i)}P_{ij} &= \begin{cases}
      -P_{Dj}, j\in \mathcal{N} \backslash \mathcal{N}_\text{IB} \\
      P_{Gj} - P_{Dj}, j\in \mathcal{N}_\text{IB}
    \end{cases} \\
    -B_{sh,i} V_i^2 + \sum_{j \in K(i)}Q_{ij}  &= \begin{cases}
      -Q_{Dj}, j\in \mathcal{N} \backslash \{\mathcal{N}_\text{IB} \cup \mathcal{N}_\text{CD}\} \\
      Q_{Gj} - Q_{Dj}, j\in \mathcal{N}_\text{IB} \\
      Q_{Cj} - Q_{Dj}, j\in \mathcal{N}_\text{CD}
    \end{cases}
  \end{aligned}
\end{equation}

The IB-ERs are typically designed with redundant rated capacity for safety reasons and operate under maximum power point tracking (MPPT) mode. Hence, the controllable range of the reactive power of IB-ERs can be determined by the rated capacity $S_{Gi}$ and maximum power output $\overline{P_{Gi}}$. The reactive power range of controllable devices is $|Q_{Gi}| \leq \sqrt{S_{Gi}^2-\overline{P_{Gi}}^2}$ and $\underline{Q_{Ci}} \leq Q_{Ci} \leq \overline{Q_{Ci}}$.

\section{SAC Algorithm}
\Cref{alg:sac} is a naive implement of SAC in \cite{haarnoja2018learning}. Several techniques to make SAC more practical and stable, such as optimization of the $\alpha$, frozen target and two Q-functions, are omitted here.
\begin{equation}
    \label{eq:bellman}
    \begin{aligned}
      Q^{\pi}(s,a) &=
      \mathop{\mathbb{E}}\limits_{s', a'}
      \Big[R(s,a,s') + \gamma \left(Q^{\pi}\left(s',a'\right) + \alpha H\left(\pi\left(\cdot|s'\right)\right)\right) \Big] \\
      & \approx r + \gamma \left( Q^{\pi}\left(s', \tilde{a}'\right) - \alpha \log \pi \left(\left. \tilde{a}'\right|s' \right)  \right), \tilde{a}'\sim \pi(\left.\cdot \right| s')
    \end{aligned}
  \end{equation}
\begin{equation}
    \label{eq:sacpi}
    \max_{\theta} \mathop{\mathbb{E}}\limits_{\substack{s\sim \mathcal{D}\\ \xi\sim \mathcal{N}}} \left[ Q\left(s,\tilde{a}_{\theta}(s,\xi)\right) - \alpha \log \pi_{\theta}\left(\left.\tilde{a}_{\theta}(s,\xi)\right|s\right) \right]
  \end{equation}
\begin{algorithm}
    \caption{Soft Actor-Critic}
    \label{alg:sac}
    Initialize experience pool, policy and value function approximators' parameter vectors\;
    \ForEach{
      episode
    }{
      \ForEach{
        environment step
      }{
        $a_t\sim\pi(\cdot|s_t)$\;
        Feed $a_t$ to the environment, get reward $r_t$ and next state $s_{t+1}$\;\;
        $\mathcal{D}\leftarrow\mathcal{D}\cup\{(s_t,a_t,r_t,s_{t+1})\}$\;
      }
      \ForEach{gradient step}{
        Update $Q^\pi$ according to \Cref{eq:bellman}\;
        Update $\pi$ according to \Cref{eq:sacpi}\;
      }
    }
  \end{algorithm}

\subsection{Simulation Environment Setup}
In the numerical simulation, steady-state distribution system reinforcement learning environments are built under the scheme of the well-known toolkit Gym\cite{brockman2016openai}. Power flow equations of IEEE 33-bus\cite{25627} and 69-bus\cite{DAS2008361}
distribution test cases are solved to realize AMDP for the offline stage and MDP for the online stage. In the 33-bus test case, three IB-ERs of 4.5 MVA are connected to node 22,25,18 and one SVC of 4 MVA is connected to node 33. In the 69-bus test case, four IB-ERs of 3.5 MVA are connected to node 31,41,6,9 and tow SVCs of 3.5 MVA are connected to node 20,60. The load and generation level are set according to scaled data of a pilot project in eastern China. In order to suggest the efficiency of the proposed two-stage algorithm and adversarial training for transfer gap, we select a stationary power flow section to get rid of data noise. The voltage limitations are set to be $[0.95, 1.05]$ in all cases. For the modeling errors (the difference between the online model and offline model), the ranges of parameters are set as $[0.5, 2.0]$.

The reward parameters are chosen to be $C_v=100$ for 33-bus case and $C_v=1000$ for 69-bus case. Note that when the power flow fails, it is treated as a failure of the system with a negative reward $r=-300$. The maximum episode length is $96$ in this study.

\section{Hyperparameters}
The hyperparameters of SAC, JASAC, ASAC and preSAC used in this paper are shown in \Cref{tbl:params}.
\begin{table}[!htbp]
  \centering
  \caption{Algorithm hyperparameters}
  \label{tbl:params}
  \begin{tabular}{@{}lll@{}}
  \toprule
  Algo. & Parameter & Value \\ \midrule
  Shared & optimizer & Adam \\
   & non-linearity & ReLU \\
   & replay buffer size & $4\times 10^{5}$ \\
   & number of hidden layers & $\{2,3\}$ \\
   & episode size & $96$ \\
   & $\eta$ & $0.995$ \\
   & $\lambda$ & $10^{-3}$ \\ \midrule
  SAC / preSAC & $\alpha$ & $\{0.07, 0.03\}$  \\ \midrule
  JASAC / ASAC & $\alpha_p$ & $\{0.07, 0.03\}$ \\
   & $\alpha_o$ & $\{0.04, 0.02\}$ \\ \bottomrule
  \end{tabular}
\end{table}

\bibliographystyle{IEEEtran}
\bibliography{paper}